\documentclass[twocolumn,pre, showpacs,amsfonts,amsmath,assymb,eufrak]{revtex4}

\usepackage{epsfig}

\usepackage{bm}% bold math
\usepackage{braket}

\usepackage{graphicx}
\newcommand{\eq}[1]{\begin{equation} #1 \end{equation}}
\newcommand{\eqa}[2]{\begin{equation} #1 \label{#2} \end{equation}}
\newcommand{\balign}[1]{\begin{align} #1 \end{align}}

\newcommand{\bs}{\boldsymbol}

\newcommand{\figin}[4]
{\begin{figure}[tb]
\centering
\includegraphics[width= #1]{#2.pdf}
\caption{#3}
\label{f:#4}
\end{figure}}

\newcommand{\todayd}{\the\year/\the\month/\the\day}

\newcommand{\bib}{\bibitem}

\newcommand{\lb}{\label}
\newcommand{\nt}{\notag}
\newcommand{\Tr}{\mathrm{Tr}}

\newcommand{\bel}{\begin{easylist}}
\newcommand{\eel}{\end{easylist}}
\newcommand{\eref}[1]{Eq.~\eqref{#1}}

\newcommand{\fref}[1]{Fig.~\ref{f:#1}}
\newcommand{\sref}[1]{Sec.~\ref{s:#1}}

\def \({\left(}
\def \){\right)}

\newcommand{\abs}[1]{\left|#1\right|}

\newcommand{\sumtwo}[2]%
{\mathop{\sum_{#1}}_{#2}}
\newcommand{\sumthree}[3]%
{\mathop{\mathop{\sum_{#1}}_{#2}}_{#3}}
\newcommand{\sumfour}[4]%
{\mathop{\mathop{\mathop{\sum_{#1}}_{#2}}_{#3}}_{#4}} 
\newcommand{\prodtwo}[2]%
{\mathop{\prod_{#1}}_{#2}}
\newcommand{\mintwo}[2]%
{\mathop{\min_{#1}}_{#2}}
\newcommand{\maxtwo}[2]%
{\mathop{\max_{#1}}_{#2}}
\newcommand{\maxthree}[3]%
{\mathop{\mathop{\max_{#1}}_{#2}}_{#3}}
\newcommand{\limtwo}[2]%
{\mathop{\lim_{#1}}_{#2}}
\newcommand{\suptwo}[2]%
{\mathop{\sup_{#1}}_{#2}}
\newcommand{\supthree}[3]%
{\mathop{\mathop{\sup_{#1}}_{#2}}_{#3}}
\newcommand{\supfour}[4]%
{\mathop{\mathop{\mathop{\sup_{#1}}_{#2}}_{#3}}_{#4}} 
\newcommand{\inftwo}[2]%
{\mathop{\inf_{#1}}_{#2}}
\newcommand{\infthree}[3]%
{\mathop{\mathop{\inf_{#1}}_{#2}}_{#3}}
\newcommand{\inffour}[4]%
{\mathop{\mathop{\mathop{\inf_{#1}}_{#2}}_{#3}}_{#4}} 
\newcommand\calD{{\cal D}}

\newcommand\calS{{\cal S}}

\newcommand{\bsn}{\boldsymbol{n}}

\newcommand{\bsx}{\boldsymbol{x}}

\newcommand{\sfP}{\mathsf{P}}

\newcommand{\bcs}{\backslash}

\def\rnum#1{\resizebox{0.5em}{\height}{\expandafter{\romannumeral #1}}}
\def\Rnum#1{\resizebox{0.5em}{\height}{\uppercase\expandafter{\romannumeral #1}}}
\newcommand{\mbra}[1]{\left\langle
\begin{matrix}
#1
\end{matrix}
\right|
}
\newcommand{\mket}[1]{\left|
\begin{matrix}
#1
\end{matrix}
\right\rangle
}

\newcommand{\deff}{D_{\rm eff}}

\newcommand{\bsgm}{{\bs \sigma}}

\begin{document}

\preprint{APS/123-QED}

\title{Analytic model of thermalization: Quantum emulation of classical cellular automata}% Force line breaks with \\

\author{Naoto Shiraishi}
\affiliation{%
Department of physics, Keio university, Hiyoshi 3-14-1, Kohoku-ku, Yokohama, Japan
}%

\date{\today}% It is always \today, today,

\begin{abstract}
We introduce a novel method of quantum emulation of a classical reversible cellular automaton.
By applying this method to a chaotic cellular automaton, the obtained quantum many-body system thermalizes while all the energy eigenstates and eigenvalues are solvable.
These explicit solutions allow us to verify the validity of some scenarios of thermalization to this system.
We find that two leading scenarios, the eigenstate thermalization hypothesis scenario and the large effective dimension scenario, do not explain thermalization in this model.

\end{abstract}


\maketitle

\section{Introduction}
Presence and absence of thermalization in quantum many-body systems is one of the most profound problems in theoretical physics.
In a broad class of quantum systems, a nonequilibrium initial state relaxes to the unique equilibrium state~\cite{KWW, Cra08, Gri, Rig08}.
This phenomenon is called {\it thermalization}.
However proverbially, some quantum systems including integrable systems and localized systems do not show thermalization~\cite{Rig08, Gri, Lan, EF}.
Thus what determines the presence or absence of thermalization and why thermalization occurs have been intensively studied.

There are mainly two approaches to tackle this problem.
One approach employs numerical simulations, which has discovered many interesting phenomena and properties in nonequilibrium relaxation dynamics~\cite{Rig08, KIH, SHP, BMH, SPR, Rig16, BCH}.
However, it is not easy to establish a theoretical framework from numerical data.
In addition, numerical simulations inevitably face the limitation of finite size and finite time, which has sometimes led to incorrect expectations~\cite{BCH, Kim}. 
The other approach investigates mathematical foundations, where many general theorems have been proven rigorously~\cite{Rei08, LPSW, Neu, Tas98, GME, SF12, FBC16, GE16, Pin, Pal}.
Nevertheless, most studies concern properties irrelevant to (non-)integrability and fail to address the difference between integrable and non-integrable systems, since it is very difficult to prove general theorems clarifying this difference.
To break this impasse, we propose another approach; constructing elaborated models which thermalize but can be treated analytically in some sense.
In many research fields, artificial but elaborated analytically-accessible models, which support or disprove some conjectures including the Haldane conjecture (Affleck-Kennedy-Lieb-Tasaki model)~\cite{AKLT}, the positive rate conjecture~\cite{Gac}, the gap decision problem~\cite{CPW}, and the entanglement area law in one-dimension~\cite{MS16}, have helped our better understanding.
The approach with analytic elaborated models will fill the disadvantages of two existing approaches, and help our further understanding of thermalization.

To explore this, we shall have a look at the field of quantum chaos, where the difference between chaotic and integrable systems  has also been a central subject~\cite{Haa}.
The universality of the energy level statistics is analytically proven in some one-body systems including quantum billiards~\cite{Ber85, SR, Mul} and quantum graphs~\cite{KS, BSW}.
The crucial step of the above analytic approach is the quantum-classical correspondence based on periodic orbits (the Gutzwiller trace formula)~\cite{Gut}, by which we can import properties of classical chaos to quantum systems.
In case of quantum billiards, for example, by assuming good chaotic properties (e.g., mixing property) in the corresponding classical billiard system~\cite{HOdA}, we obtain detailed results on the quantum chaotic billiard.
Unfortunately, such a quantum-classical correspondence has not yet been established for many-body systems.

In this paper, we propose a novel type of quantum model which emulates (quantizes) a classical cellular automaton (CA).
This model is shift-invariant, has local interaction, and its local Hilbert space is small. 
These properties are conventionally required for physical many-body systems.
The emulation can be performed for both integrable and chaotic CA.
The advantage of this model lies in the fact that all the energy eigenstates and eigenenergies are solvable even in case of a chaotic CA.
Using the expression of energy eigenstates, we verify two leading scenarios of thermalization~\cite{Tas16} analytically.
Maybe surprisingly, although this model thermalizes, this model neither satisfy the above two scenarios.

This paper is organized as follows.
Before discussing quantum systems, in \sref{RCA}, we briefly explain the classical second-order reversible CA.
In \sref{qmodel}, we construct the quantum model emulating the classical second-order reversible CA.
Notably, all the energy eigenstates and eigenenergies of this quantum model are formally solvable regardless of whether the emulated classical CA is integrable or chaotic.
This fact is demonstrated in \sref{solve}.
With the help of this formal solution, in \sref{analytic}, we analytically show that this quantum model thermalizes but this thermalization is not explained by two leading scenarios of thermalization.
In \sref{gen}, we generalize this model in order to recover the extensivity with keeping analytic properties.

\figin{9cm}{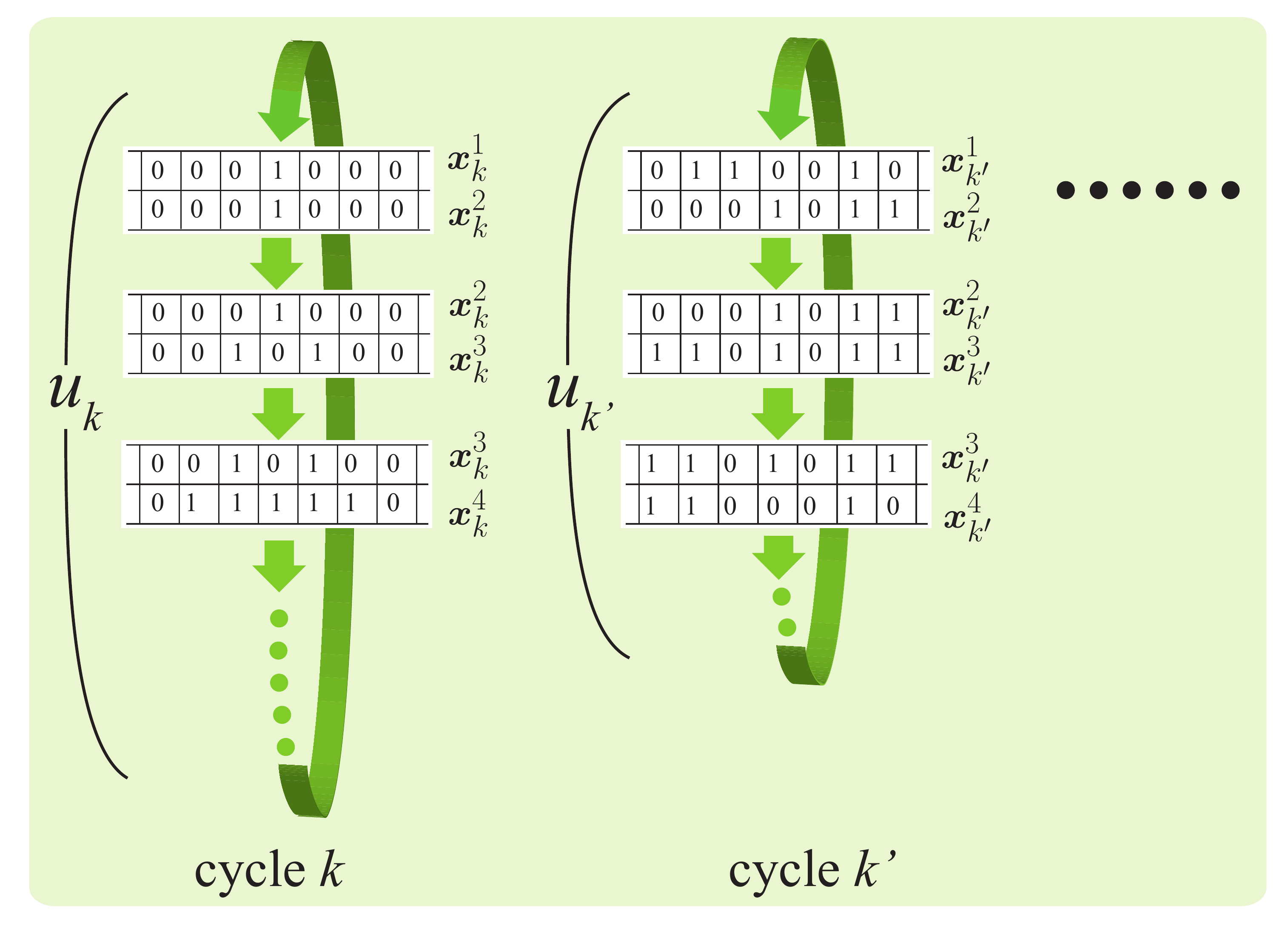}{
(Color online) 
Schematic of the length of cycles of a second-order reversible CA with the rule 214R.
In such a CA, the (generalized) state of the system is given by a pair of the present state and the state one-step before; $(\bsx ^n_k, \bsx ^{n+1}_k)$.
The length of the cycle $k$ is denoted by $u_k$.
}{rule}

\section{Second-order reversible cellular automata}\lb{s:RCA}
Before constructing the model, we briefly review reversible classical CA.
We first explain a standard one-dimensional CA, which is not reversible in general, with length $L$ with the periodic boundary condition.
Suppose that each site takes $d$ possible states in $\calS_d:=\{ 0,1,\cdots , d-1\}$.
For simplicity, we consider the case that the state of the site $i$ at the next step depends only on the present states of sites $i$ and $i\pm 1$.
By denoting by $x_i^n$ the state of the site $i$ at the $n$-th step, the rule of a CA is expressed as $x_i^{n+1}=f(x_{i-1}^n, x_i^n, x_{i+1}^n)$.
Here, $f$ is a map of $(\calS_d) ^3\to \calS_d$ and thus there are $d^{(d^3)}$ possible rules.
In particular, the CA with $d=2$ has 256 rules, which are labeled from 0 to 255 by Wolfram code~\cite{Wol}.
The explicit definition and an example of Wolfram code is presented in Appendix.\ref{a-Wcode}.

Using the map $f$, we construct a kind of reversible CA named {\it second-order reversible CA} as
\eq{
x_i^{n+1}\equiv f(x_{i-1}^n, x_i^n, x_{i+1}^n)-x_i^{n-1}  \mod d.
}
It is easy to check that if a trajectory of time evolution $\cdots \to \bsx ^{n-1}\to \bsx^n\to \bsx ^{n+1}\cdots$ can realize under a rule, then its time reversal $\cdots \to \bsx ^{n+1}\to \bsx^n\to \bsx ^{n-1}\cdots$ can also realize under the same rule.
By regarding the pair of a present state $\bsx^n$ and its last-minute state $\bsx^{n-1}$ as a (generalized) state, the time evolution $(\bsx^{n-1}, \bsx^n)\to (\bsx^n,\bsx^{n+1})$ becomes a one-to-one map of states.
In this picture, the state space has $d^{2L}$ possible states.
The rules of such reversible CA with $d=2$ are labeled by adding R to the corresponding Wolfram code such as 214R. 
Because the number of possible states is finite, the state space is decomposed into some cyclic trajectories of time-evolution, which we call {\it cycle} (see \fref{rule}).
We label a cycle as $k$ ($1\leq k\leq K$), where $K$ is the number of all possible cycles.
We denote the length of a cycle $k$ by $u_k$, which satisfies $\sum_k u_k=d^{2L}$.
For each cycle $k$, we fix a state as the initial state $\bsx_k^1$, and write the state of the site $i$ at the $n$-th step as $x_{k,i}^n$.

We here summarize some known properties of reversible CA with $d=2$.
Takesue~\cite{Tak, Takp} has reported its thermodynamic properties.
Some CA (e.g., 90R) have local conserved quantities, which can be regarded as a counterpart of integrable systems.
Some CA (e.g., 73R) have localized states, i.e., if a certain local structure appears in the initial state, this structure never disappears through time evolution.
Other CA (e.g., 214R) neither have local conserved quantities nor localized states, and some of them show chaotic behavior in numerical simulations.
An example is seen in Ref.~\cite{Wol}, where the CA with the rule 214R appears to thermalize.
Throughout this letter we use the word {\it chaotic CA} in a loose sense that the CA thermalizes (see also Appendix.\ref{a-term}).

The length of the cycles has also been numerically investigated~\cite{Tak, Takp}.
In some CA the maximum and averaged length of cycles is exponentially large ($u\sim 2^L$).
However, in any case its proportion to the size of the state space is exponentially small ($u/d^{2L}\sim 2^{-L}$).
We note that in any rule there exists a cycle with very short length, which stems from the fact that any CA keeps spatial periodicity~\cite{period}.
%We now consider the cycle that such a state belongs to.
%Because the number of states with spatial period $r$ is only $(2d)^r$, the length of the cycle is less than $(2d)^r$.

It is worth comparing the chaotic CA and the chaotic classical billiard system.
Similar to the chaotic billiard system, the chaotic CA has many periodic orbits.
In contrast, differently from the chaotic billiard system, the chaotic CA shows the mixing property only locally, not globally.
This difference is considered to come from the difference between few-body chaos and many-body chaos.

\section{Quantum emulation of classical CA}\lb{s:qmodel}
We now introduce a quantum system which emulates a second-order classical reversible CA.
Consider a three-layered one-dimensional quantum system with length $L$, where the top row is {\it head system} and the bottom two rows are  {\it CA system}.
The head system consists of a single free fermion, which controls the dynamics of the CA system.
The CA system consists of $2L$ spins with $d$ degrees of freedom, which emulates a given classical CA.

\figin{8cm}{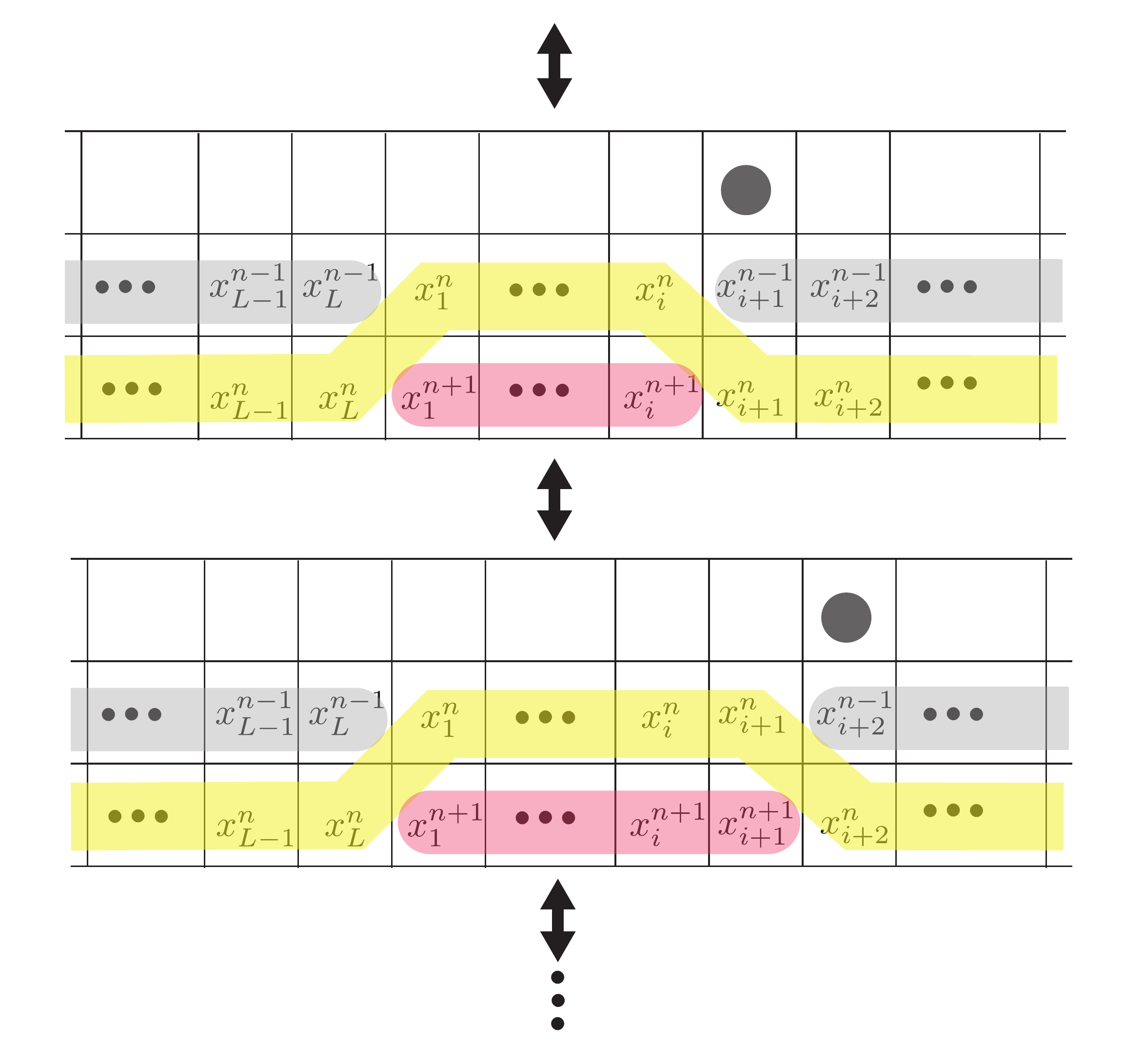}{
(Color online)
Schematic of how the quantum system emulates the classical CA.
The top row and the below two rows correspond to the head system and the CA system, respectively.
The gray, yellow, and red thick lines represent the state of the classical CA at $n-1$, $n$, and $n+1$ step, respectively.
By applying the Hamiltonian $H$ to the system, the fermion moves right and the sites of the CA system just under the fermion evolves one step (and its time-reversal transition occurs).
}{qrule}

Let $a_{1,i}$ and $a_{2,i}$ be the states of the $i$-th site in the first and second layer of the CA system, and $c_i$ and $c_i^\dagger$ be the annihilation and creation operator of the head fermion at the $i$-th site.
By denoting the product state of the CA system at sites $i$ and $i\pm 1$ as
\eq{
\mket{
a_{1,i-1} &a_{1,i} & a_{1,i+1} \\
a_{2,i-1} & a_{2,i} & a_{2,i+1}
}_i, \nt
}
the local Hamiltonian is expressed as (see \fref{qrule}) %care
\eq{
h_i=\sum_{p,q,r,s \in \calS_d}
\mket{
p &r& * \\
* &X &s
}
\mbra{
p &q & * \\
* &r &s
}_i
\otimes c^\dagger_{i+1}c_i +{\rm c.c.}
}
for $2\leq i\leq L-1$, where $X$ is determined by using the rule of the CA as 
\eq{
X\equiv f(p,r,s)-q \mod d. 
}
The symbol $*$ in the bracket means that these bra and ket do not operate on this site.
This local Hamiltonian represents update of the $i$-th site of the CA system and shift of the fermion to the next site.
Its complex conjugate represents the backward process of above.
The boundary condition is set to
\balign{
h_L&=\sum_{p,q,r,s \in \calS_d}
\mket{
p &r& s \\
* &X &*
}
\mbra{
p &q & s \\
* &r &*
}_L
\otimes c^\dagger_{1}c_L +{\rm c.c.},
\\
h_1&=\sum_{p,q,r,s \in \calS_d}
\mket{
* &r& * \\
p &X &s
}
\mbra{
* &q & * \\
p &r &s
}_1
\otimes c^\dagger_{2}c_1 +{\rm c.c.}.
}
The total Hamiltonian of the system is given by $H=\sum_{i=1}^L h_i$.
Owing to the boundary condition, a single circle move of the fermion $1\to 2\to \cdots \to L\to 1$ induces one-step time evolution of the CA.

\section{Exact energy eigenstates and eigenvalues}\lb{s:solve}
A remarkable point of this model is that all the energy eigenstates and eigenvalues can be explicitly written down with the help of the knowledge of the emulated classical CA.
We introduce a basis of the CA system written as
\eq{
\ket{X_{k,i}^n}:=\mket{
x_{k,1}^n & \cdots &x_{k,i-1}^n &x_{k,i}^{n-1}&\cdots x_{k,L}^{n-1} \\
x_{k,1}^{n+1} & \cdots &x_{k,i-1}^{n+1} &x_{k,i}^n&\cdots x_{k,L}^n
}
}
with $1\leq n\leq u_k$, $1\leq i\leq L$, and $1\leq k\leq K$, which we also call {\it computational basis}.
Using this, all the energy eigenstates and corresponding eigenenergies are expressed as
\balign{
\ket{E_{k, m}}=&\frac{1}{\sqrt{u_kL}}\sum_{n=1}^{u_k} \sum_{i=1}^L e^{-\frac{2\pi {\rm i} m (nL+i)}{u_kL}} \ket{X_{k,i}^n}\otimes \ket{i} \lb{eigenstate} \\
E_{k,m}=&2\cos \frac{2\pi m}{u_kL} \lb{eigenenergy}
}
with $m=0,1,\cdots u_kL-1$.
Here, $\ket{i}$ represents the state of the head system that the fermion is at the site $i$.
The form of the energy eigenstate directly follows from a simple but crucial relation 
\eq{
H (\ket{X_{k,i}^n}\otimes \ket{i})=\ket{X_{k,i+1}^n}\otimes\ket{i+1}+\ket{X_{k,i-1}^n}\otimes\ket{i-1}. 
}
The structure of the solution is close to that of a free fermion, while the length in the state space is elongated from $L$ to $u_kL$ (see \fref{solve}).
We note that some CA has $u_k=O(e^L)$.

\figin{8cm}{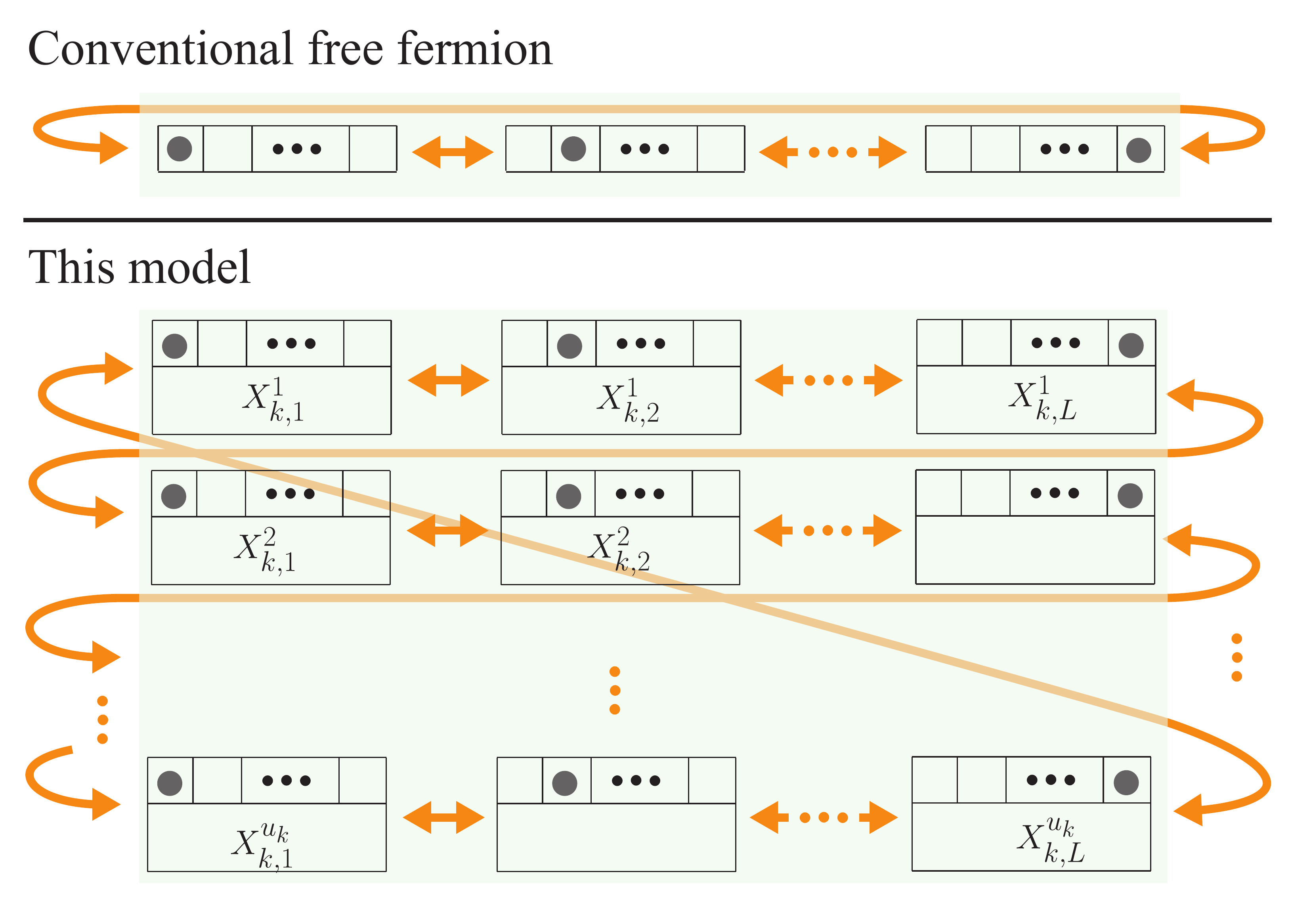}{
(Color online)
The state space of the conventional (single) free fermion system and that of this model.
Although the structure of these two is similar, the length of the cyclic path of this model is elongated from $L$ to $u_kL$.
The factor $u_k$ might increase exponentially with respect to $L$.
}{solve}

\section{Analytic results on quantum thermalization}\lb{s:analytic}
Our calculation on the quantum model before here does not rely on properties of emulated CA.
To investigate thermalization phenomena by using this model, we now focus on chaotic CA.
Although we do not specify the concrete rule, on the basis of the numerical observations~\cite{Wol, Tak, Takp}, it is highly plausible that some CA with some $d$ is indeed a chaotic CA.
In the track of quantization of few-body billiard systems, we {\it assume} the existence of a second-order reversible chaotic CA, which satisfies the following three properties in the thermodynamic limit (The precise statements are shown in Appendix.\ref{a-state-cond}):
\begin{enumerate}
\item {\it Thermalization}:
If we observe only a local region ${\rm C}_{1,l}$ with fixed $l$, then time evolution from an initial state with no spatial periodicity provides the uniform distribution of possible $d^{2l}$ states.
\item {\it Many cycles}:
The maximum length of a cycle is exponentially small compared to the number of possible states; $d^{2L}$.
\item {\it No coherence}:
If we observe a region ${\rm C}_{l+1,L}$ with fixed $l$, then almost all states appear at most once in a fixed cycle.
\end{enumerate}
We here denoted by ${\rm C}_{i_1,i_2}$ ($i_1<i_2$) a subsystem of the CA system with sites $i_1\leq i\leq i_2$.
(If $i_2>L$, then ${\rm C}_{i_1,i_2}$ represents the subsystem with sites $i_1\leq i\leq L$ and $1\leq i\leq i_2-L$).
The counterparts of the conditions (i) and (ii) in a chaotic billiard system are the mixing property and the fact that there are exponentially many periodic orbits.
The condition (iii) is strongly suggested by the condition (ii) for the following surmise:
The number of states which share states of all sites except $C_{1,l}$ is only $d^{2l}=O(1)$, while a single cycle covers exponentially small proportion of the exponentially large state space.
Thus, it is highly plausible to consider that such a short cycle passes a set of states with the size $O(1)$ at most once.

We fix a chaotic CA which satisfies the aforementioned conditions.
Then, in terms of macroscopic observables of the CA system, all the energy eigenstates without spatial periodicity are thermal.
%in the sense that the partial trace to a small subspace with finite size, $\Tr _{{\rm H}, {\rm C}_{l+1,L}}[\ket{E}\bra{E}]$, is equal to the Gibbs state of ${\rm C}_{1,l}$.
This fact is guaranteed by the condition (i) and (iii):
The condition (i) ensures the equipartition in view of the computational basis, and the condition (iii) ensures the absence of coherence.
Since we cannot prepare a truly spatially-periodic state in a macroscopic system at finite temperature, we confirm that the CA system thermalizes after a physical quench~\cite{comm-therm}.

Owing to the exact solutions, we draw many analytic results on this model.
We in particular verify the validity of two leading scenarios of thermalization, the eigenstate thermalization hypothesis (ETH) scenario and the large effective dimension scenario~\cite{Tas16}, in this model.
The first scenario relies on the ETH, which claims that all the energy eigenstates are thermal~\cite{Neu, Deu, Sre, HZB, Tas98, Rig08}.
The ETH is known to be a sufficient condition for thermalization~\cite{GE16, Tas16}.
Numerical simulations show that the ETH is indeed satisfied in many non-integrable models~\cite{Rig08, KIH, SHP, BMH}, and thus the ETH is believed to be satisfied in chaotic thermalizing systems.
Contrary to this, the ETH with respect to macroscopic observables in the CA system is not satisfied in our model.
The violation of the ETH stems from the fact that spatially periodic states have very short period as explained and the corresponding energy eigenstates are not thermal.
This model is another counterexample to the ETH different from Refs.~\cite{SM, MS}.
We remark that the violation of the ETH is inherent to the emulation of a CA and it does not rely on the assumptions (i)-(iii).

The second scenario of thermalization is the large effective dimension scenario~\cite{LPSW, GHH, Tas16}, which claims that an initial state not concentrated on small number of energy eigenstates thermalizes.
The effective dimension $\deff$ of a pure state $\ket{\psi}$ is defined as 
\eq{
\deff :=\( \sum_n \abs{\braket{E_n|\psi}}^4\) ^{-1}, 
}
where $\ket{E_n}$ is the $n$-th  energy eigenstate.
The effective dimension takes $1\leq \deff \leq D$ with the dimension of the Hilbert space of the energy shell $D$, and it quantifies how many energy eigenstates the state $\ket{\psi}$ effectively covers.
It has been proven that if the effective dimension of an initial state is not exponentially small compared to $D$ (i.e., $\deff/D={\rm poly}(1/L)$), then this initial state thermalizes~\cite{Tas16, comm-deff}.
The precise statement of this theorem is shown in Appendix.\ref{a-Deff}. 
Numerical simulations on some specific models support the large effective dimension scenario~\cite{SPR, Rig16}.
In our model, the effective dimension of some initial states can be calculated explicitly.
Let us take an initial state $\ket{\psi_{\rm ini}}$ such that the head system is $\ket{i}$ for some $i$ and the CA system is one of the computational basis vectors.
Then, the effective dimension of $\ket{\psi_{\rm ini}}$ is exactly same as the length of the cycle to which the state belongs, and the condition (ii) says that it is exponentially small compared to the dimension; $D=(2d)^L$.
It is hence concluded that thermalization in a sector with the aforementioned initial state is not explained by the large effective dimension scenario.

\section{Generalization to many head particles}\lb{s:gen}

\subsection{Model construction}

\figin{4cm}{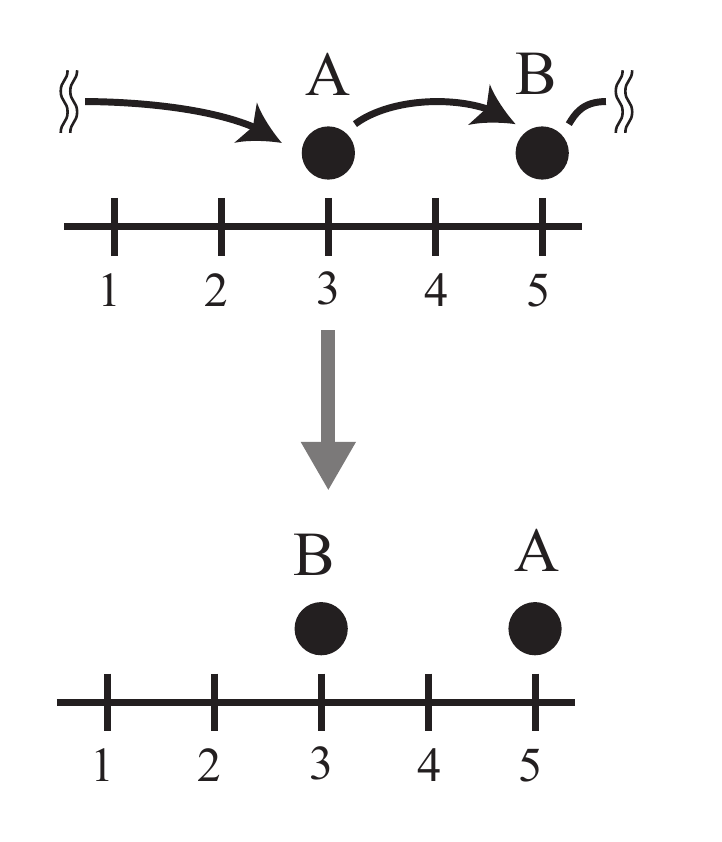}{
An example of a single circle move in case of $L=5$ and $M=2$.
In a circle move, the fermion A moves as $3\to 4\to 5$ and B moves as $5\to 1\to 2\to 3$.
}{circle-move}

The presented model has only a single fermion in the head system and thus $L\to\infty$ limit is not the conventional thermodynamic limit.
To realize the conventional thermodynamic limit, we generalize our model to many fermions (or hard-core bosons) with a little modified assumptions on CA.

Replace the head system from a single fermion to $N$ interacting fermions (or hard-core bosons) whose hopping still couples to update of the CA system.
We suppose that these particles jump only to its neighboring sites.
The Hamiltonian with length $L$ and $N$ particles is then constructed as
\balign{
H=&\sum_{i=1}^L h_i +V(\bsn ) \\
h_i=&\sum_{p,q,r,s \in \calS_d}
\mket{
p &r& * \\
* &X &s
}
\mbra{
p &q & * \\
* &r &s
}_i
\otimes t_i c^\dagger_{i+1}c_i +{\rm c.c.} ,
}
where we defined $\bsn :=\{ n_1,n_2,\ldots ,n_L\}$, and $n_i:=c_i^\dagger c_i$ is the number operator at the $i$-th site.
The functional form of $V$ is arbitrary, and the coefficient of hopping $t_i$ is position-dependent in general.
The Hamiltonian of the head system
\eqa{
H_{\rm h}=\sum_i  t_i c^\dagger_{i+1}c_i +{\rm c.c.} + V(\bsn )
}{gen-head-H}
can be either integrable or chaotic.
We remark that differently from the case with a single fermion the boundary condition is set as the periodic boundary condition.

\figin{8cm}{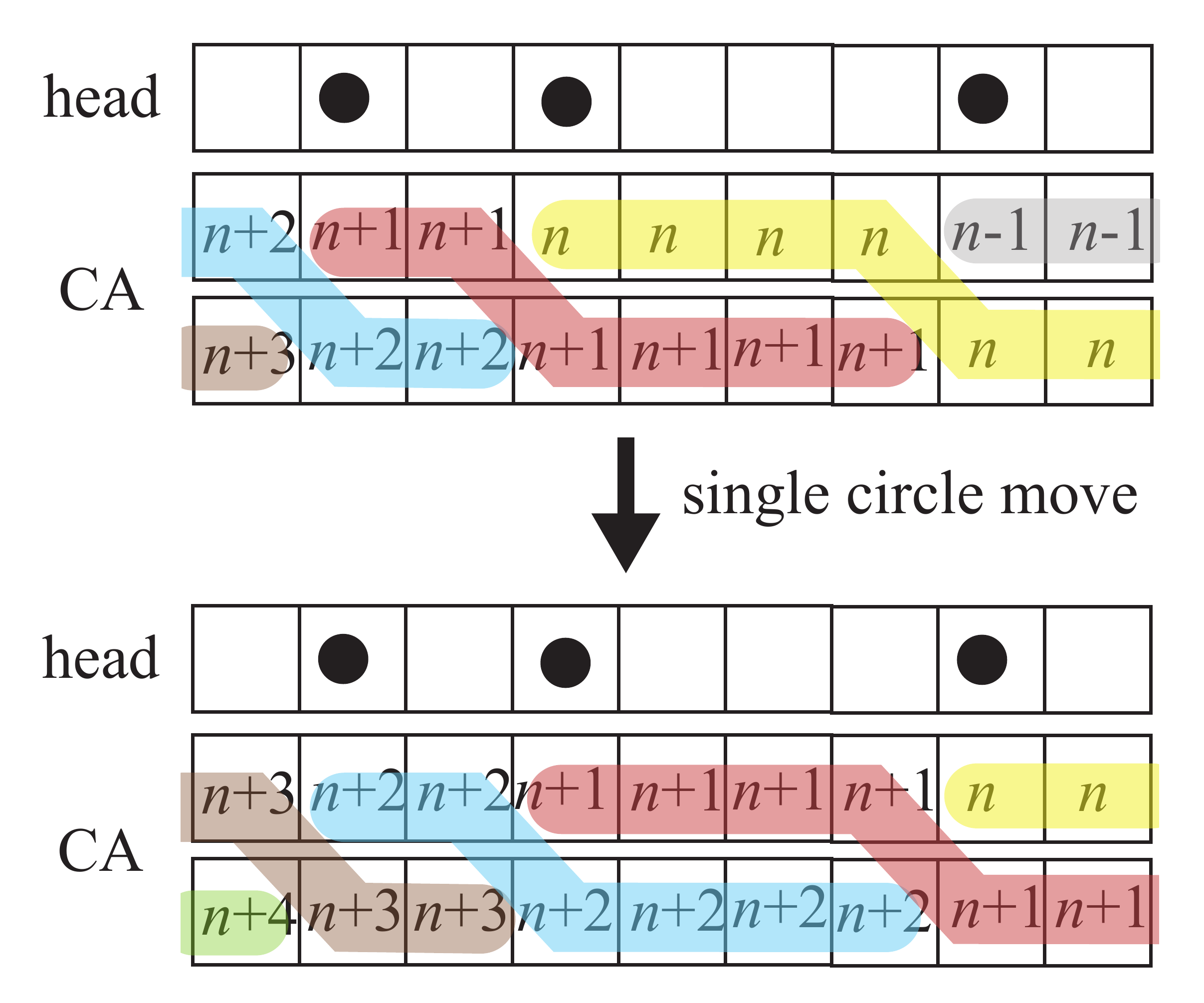}{
Schematic of an example of the quantum system for $L=9$ and $N=3$.
The number in sites in the CA system represents the present step of the state of the corresponding classical modified CA.
With a single circle move, the state of the CA system is updated by one step.
}{many-head}

In this system, a {\it circle move} of head particles represents the move of head particles that each particle moves to the position of the nearest rightmost particle in the present configuration.
Take a case with $L=5$ and $N=2$ as an example (see \fref{circle-move}).
We label two fermions as A and B for convenience of explanation with keeping in mind that these two are in fact indistinguishable.
Suppose that the fermion A is at the site 3, and the fermion B is at the site 5.
The circle move of these head particles means the move of A as $3\to 4\to 5$ and B as $5\to 1\to 2\to 3$.
Then, a single circle move of head particles updates the state of the CA system by one step (See also \fref{many-head}, which depicts the case of $L=9$ and $N=3$).
Remark that the rule of the classical CA emulated by this model is slightly different from the conventional one in that the update at the boundary refers states of $N$-step before and after.
We call this classical CA as {\it modified CA}.
We emphasize that this modified CA is still a classical CA, and to obtain the solution of this modified CA we need not to solve quantum fermion problems.
Under this modified CA rule, $4^L$ possible states of the CA are decomposed into cycles, and we safely define the length of a cycle.
Thus, by fixing a basing configuration of head particles and the initial state of the CA system on the $k$-th cycle, the state of the CA system at the $n$-th step ($0\leq n\leq u_k-1$) with the present configuration of the head system $\bsgm$ is uniquely determined, which we denote by $\ket{X^n_{k,\bsgm}}$.

Suppose that the state of the CA system belongs to the cycle $k$.
Then, $u_k$-times circle moves of fermions in the head system convey the total system back to the original state.
To reflect this fact, we depict as an example the state space in case of $L=5$, $N=2$, and $u_k=3$ in \fref{many-head-space}.
In this figure, we omit the details of the state of the CA system.
The state space of only the head system (\fref{many-head-space}.(a)) is extended  by $u_k$ times, in which we can see the similarity to the case with a single fermion shown in \fref{solve}.

\figin{9cm}{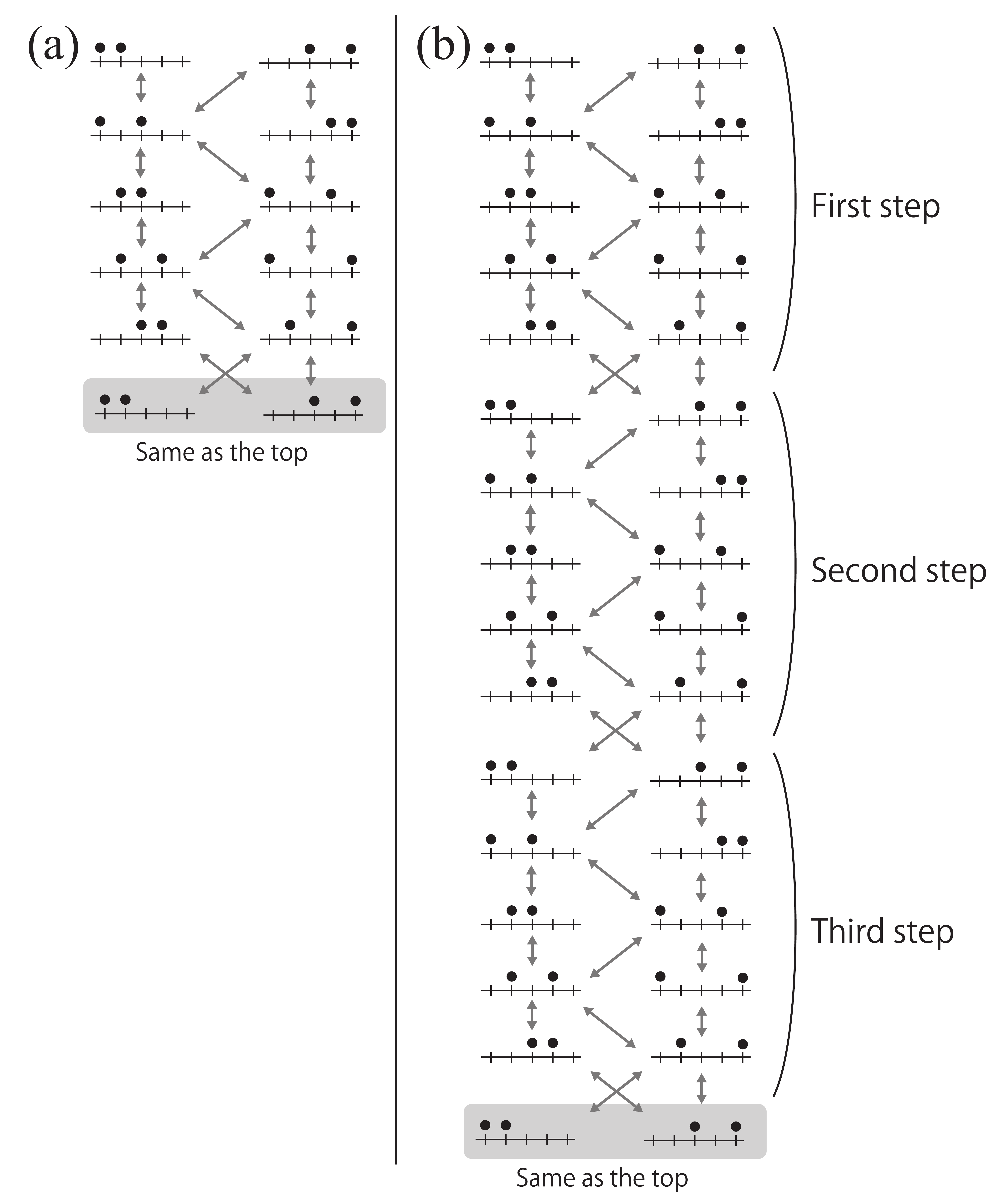}{
(a) State space of the head system with $L=5$ and $N=2$ without the CA system.
(b) State space of the total system with $L=5$, $N=2$, and $u_k=3$, where we dropped the details of the CA system.
The state space is elongated three (=$u_k$) times from that in (a), which is completely parallel to the case with a single fermion shown in \fref{solve} in the main text.
}{many-head-space}

To obtain the energy eigenstates of the total system $H$, we consider $H_{\rm h}$, the Hamiltonian of only the head system \eqref{gen-head-H}, with $2\pi s$ phase-twisted boundary condition.
This boundary condition means that if a particle hops from the site $L$ to 1, then the phase $e^{2\pi s i}$ is multiplied to the state.
We denote by $\ket{\Phi_j(s)}$ the $j$-th energy eigenstate ($1\leq j\leq a:={\tiny \begin{pmatrix}L \\ N\end{pmatrix}}$) of $H_{\rm h}$ with $2\pi s$ phase-twisted boundary condition.
We also denote by $E_j(s)$ the corresponding eigenenergy.
We expand $\ket{\Phi_j(s)}$ with spatial configurations as
\eq{
\ket{\Phi_j(s)}=\sum_\bsgm c_\bsgm^{j}(s)\ket{\bsgm},
}
where $\bsgm$ represents $a={\tiny \begin{pmatrix}L \\ N\end{pmatrix}}$ possible configurations of $N$ indistinguishable particles on $L$ sites.

Using these symbols, the energy eigenstate of the total system is explicitly written down as
\eqa{
\ket{j,p}:=\frac{1}{\sqrt{au_k}}\sum_{n,\bsgm}e^{2\pi i s(p) n}  c_\bsgm^{j}\(  s(p) \) \ket{\eta_\bsgm}\otimes \ket{X_{k,\bsgm}^n},
}{gen-eigen}
where $n$ runs $0\leq n\leq u_k-1$, $j$ takes $j=1,2,\cdots ,a$, and
\eq{
s(p)=\frac{p}{u_k}
} 
with $p=0,1,\cdots, u_k-1$.
The eigenenergy of $\ket{j,p}$ is given by $E_j(p/u_k)$.
We remark that the Hamiltonian of the head system $H_{\rm h}$ can be a chaotic Hamiltonian.
In this case, we cannot calculate $\ket{\Phi_j(s)}$ explicitly, while the formal solution of the total system is still given by \eref{gen-eigen}, which is still helpful to discuss some properties of thermalization.

The trick of the eigenstates with the phase-twisted boundary condition have already been seen in the case with a single fermion in \sref{analytic}.
In this case, energy eiogenstates and eigenenergies of the head system with no phase-twist is written as
\balign{
\ket{E_{l}}=&\frac{1}{\sqrt{L}} \sum_{i=1}^L e^{-\frac{2\pi {\rm i} li }{L}}\ket{i} \\
E_{l}=&2\cos \frac{2\pi l}{L}
}
with $l=0,1\ldots ,L-1$, which correspond to the case of $m=0, u_k, 2u_k, \cdots, (L-1)u_k$ in Eqs.~\eqref{eigenstate} and \eqref{eigenenergy}.
Other solutions with other $m$s correspond to the solutions of the head system with phase-twist $2\pi i/u_k, 4 \pi i/u_k,\cdots , 2(u_k-1)\pi /u_k$.
The solutions with $m=1, u_k+1, 2u_k+1, \cdots, (L-1)u_k+1$ of \eref{eigenstate}, for example, correspond to energy eigenstates of the head system with $2\pi i/u_k$ phase-twisted boundary condition .
The reason why phase-twist appears is as follows.
A state of the head system goes back to the original state with a single circle move of fermions.
In contrast, if the CA system is attached to the head system, the total state goes back to the original state with $u_k$ circle moves, and a single circle move not necessarily conveys the state to the original one.
Hence, there is additional arbitrariness of the phase $2\pi i/u_k, 4 \pi i/u_k,\cdots , 2(u_k-1)\pi i/u_k$ per single circle move.
This is the origin of the phase-twisted boundary condition.

\subsection{Analytic result of thermalization in this generalized model}

To construct a model of thermalization, we put assumptions on the modified classical CA as we do in \sref{analytic}.
Since the time evolution of the bulk of the modified CA is same as that of the conventional reversible CA, we assume that there exists a chaotic modified CA which satisfies the three conditions (thermalization, many cycles, no coherence) presented in \sref{analytic}.
In the following, we consider the emulation of this CA.

We now investigate some analytic properties of this quantum model.
First, if the initial state of the CA system is not spatially-periodic in the modified sense (i.e., the state is spatially-periodic in the bulk and the boundary condition connecting $N$-step before keeps this periodicity), this system thermalizes.
If the Hamiltonian of the head system $H_{\rm h}$ is integrable, thermalization occurs only in the CA system as in the case of a single fermion.
In contrast, if $H_{\rm h}$ is chaotic, thermalization occurs not only in the CA system but in the total system.

Next, we consider the validity of the ETH.
We first take a cycle of a classical conventional CA (not the modified CA) with the spatial period $r$.
Then the length of this cycle must divide $(d^{2r})!$ because the number of possible states with the spatial period $r$ is $d^{2r}$.
Since the particle number diverges in the thermodynamic limit, we safely assume that $N$ is a multiple of $(d^{2r})!$, which leads to $x_1^n=x_1^{n+N}$, $x_L^n=x_L^{n+N}$ for any $n$ in the classical conventional CA with the spatial period $r$.
This directly implies the crucial fact that as for this cycle the conventional CA and the modified CA are completely the same.
Hence the corresponding energy eigenstate of the quantum system has the spatial period $r$, which is not a thermal energy eigenstate.
We thus conclude that this quantum system does not satisfy the ETH.

We can also evaluate the effective dimension of some initial states.
The dimension of the Hilbert space of the total system is $D=D_{\rm head}\cdot 2^{2L}$, where $D_{\rm head}$ is the dimension of the Hilbert space of the head system in the energy shell with the corresponding energy.
We now calculate the effective dimension of the state $\ket{\bsgm}\ket{X_k}$, where $\ket{X_k}$ is a state in the computational basis on the cycle $k$.
The functional form of the energy eigenstates \eqref{gen-eigen} suggests that the effective dimension of this state is bounded above by $\deff \leq D_{\rm head}u_k$.
Since $u_k/2^{2L}$ is exponentially small with respect to $L$, we find that $\deff /D$ is exponentially small, and thus the thermalization of this model is not explained by the large effective dimension scenario.

\section{Discussion}
We have introduced a quantum model that emulates a classical reversible CA.
Differently from existing ideas of quantum CA~\cite{QCA, Werner} and quantum emulation of classical computation~\cite{Fey}, our model achieves emulation of stationary dynamics with a local and static Hamiltonian.
With the help of the knowledge on the emulated classical CA, we can fully solve its energy eigenstates and eigenenergies, which gives a great advantage to our model.
The level statistics of this model, for example, can be explicitly written down.
In particular, emulation of a chaotic CA provides a solvable model of thermalization, which serves as a good stage to examine some existing scenarios of thermalization.
Maybe surprisingly, although our model thermalizes, this thermalization cannot be explained by two leading scenarios, the ETH scenario and the large effective dimension scenario.

The violation of the ETH and thermalization coexist because we cannot sample non-thermal energy eigenstates in preparable initial states.
Although completely spatially-periodic states do not thermalize, a single {\it defect} which destroys spatial periodicity is sufficient to induce thermalization and thermal noise inevitably causes defects.
Essentially the same point has already been discussed in Refs.~\cite{SM, MS}.
This shows clear contrast to integrable systems where non-thermal energy eigenstates have negligibly small fraction, while physically plausible initial states can have exponentially heavy weight on these non-thermal energy eigenstates~\cite{Bir}.

\begin{acknowledgements}
The author thanks Shinji Takesue for helpful advice on the reversible CA and Keisuke Fujii for fruitful discussion on quantum computation.
The author also thanks Eiki Iyoda and Hiroyasu Tajima for useful comments.
This work was supported by Grant-in-Aid for JSPS Fellows JP17J00393.

\end{acknowledgements}

\appendix

\section{Clarification of terminology}\lb{a-term}

We here clarify some of terminology used in this paper.
In case that a term does not have an undisputed definition and characterization, we put only some explanations on it.

\

{\bf Integrable/non-integrable}:
Although there is no undisputed definition of quantum (non-)integrability, in this paper we call a system {\it non-integrable} if the system has no local conserved quantity.
If a system has some local conserved quantity, we call this system {\it integrable}.
In this definition, an integrable system is not necessarily an {\it exactly solvable} system, which has sufficiently many local conserved quantities to determine each energy eigenstate.
We note that a recent study~\cite{S18} succeeds in proving the non-integrability in a specific model, the XYZ chain with a magnetic field.

\

{\bf Thermal state}:
To give a precise definition of the thermal state in a macroscopic system, we first introduce a macroscopic observable (in a one-dimensional system).
An observable $A$ is called {\it macroscopic observable} if $A$ is a sum of local observables $A=\sum_i A_i$ (i.e., the support of $A_i$ is contained by $[i-r, i+r]$ with a fixed constant $r$).
Then, a state $\ket{\Psi}$ of a system X is {\it thermal} if any macroscopic observable $A$ satisfies
\eq{
\lim_{L\to \infty}\bra{\Psi}A\ket{\Psi}=\lim_{L\to \infty}\Tr[A\rho^{\rm mc}_{X}],
}
where $\rho^{\rm mc}_{X}$ is a microcanonical ensemble of $X$ with energy $\bra{\Psi}H\ket{\Psi}$.
If a state $\ket{\Psi}$ is thermal, then the partial trace to any subsystem with finite size $X'\subset X$ turns to be the Gibbs state of this subsystem:
\eq{
\lim_{L\to \infty}\Tr _{X\bcs X'}[\ket{\Psi}\bra{\Psi}]=\lim_{L\to \infty}\Tr _{X\bcs X'}[\rho^{\rm mc}_{X}].
}

\

{\bf Chaos}:
We here elucidate a sharp difference between the case of few-body systems and many-body systems.
In case of few-body systems, we can measure any observable of the system.
The notion of few-body chaos is characterized on the basis of this fact.
{\it A classical few-body system is chaotic} if for almost all initial states the long-time average of any Lebesgue measurable observable is equal to its ensemble average of microcanonical ensemble (i.e., ergodicity in the phase space).
{\it A quantum few-body system is chaotic} if for all initial states the expectation value of any Lebesgue measurable observable after relaxation is equal to its ensemble average of microcanonical ensemble.
In the quantum case, we keep $\hbar \to 0$ limit (semiclassical limit) in mind.
We remark that (1) classical systems allow exceptional initial states with measure zero, (2) classical systems needs long-time average.

By contrast, in case of many-body systems, we can measure only macroscopic observables, not all observables.
Therefore, the term {\it thermal} is defined with respect to macroscopic observables.
In a similar manner, {\it a classical many-body system is chaotic} if for almost every initial state the long-time average of any macroscopic observable is equal to its ensemble average of microcanonical ensemble.
{\it A quantum many-body system is chaotic} if for all initial states the expectation value of any macroscopic observable after relaxation is equal to its ensemble average of microcanonical ensemble.
We emphasize that we do not require ergodicity in the phase space.
In fact, no classical CA and no quantum many-body system show ergodicity in this sense.
In many-body systems, the restriction of observable is crucial for characterization of thermalization and chaos.

We note that the Wigner-Dyson level statistics and other connections to random matrix are NOT the definition of chaos, but frequently-appearing properties in quantum chaotic systems.

\

{\bf Eigenstate thermalization hypothesis (ETH)}:
We call that a system (a Hamiltonian) satisfies the {\it ETH} if all the energy eigenstates are thermal in the aforementioned sense.
For specialists, we remark that we use the word ETH in the sense of the {\it diagonal ETH}, and we do not care about the {\it off-diagonal ETH} in this paper.
We distinguish the diagonal ETH and the off-diagonal ETH because these two plays different roles in the context of thermalization.
The diagonal ETH confirms that the long-time average is equal to the microcanonical ensemble, and the off-diagonal ETH confirms that the time-series fluctuation is small.
However, diverging effective dimension also confirms that the time-series fluctuation is small~\cite{Rei08, SF12}.
Related technical points on the ETH is seen in Ref.~\cite{SMreply}

%%%%%%%%%%%%%%%%%%%%%%%%%%%%%%%%%%%%%%%
\section{Wolfram code of cellular automata with $d=2$}\lb{a-Wcode}

We here describe the Wolfram code of CA with $d=2$, which makes correspondence between the function $f$ and an integer $z\in \{ 0,1,\cdots ,255\}$~\cite{Wol}.
The correspondence is given by
\eq{
z=\sum_{a,b,c}f(a,b,c)2^{4a+2b+c},
}
where $a$, $b$, and $c$ take 0 or 1.
For example, the rule 214 ($=128+64+16+8+4$) describes the following transition rule:
\begin{table}[h]
%\caption{Table of the rule 214R}
\lb{rule}
  \begin{tabular}{ c c c c c c c c  } \hline
111 & 110 & 101 & 100 & 011 & 010 & 001 & 000  \\ \hline
1 & 1 & 0 & 1 & 1 & 1 & 0 & 0  \\ \hline
  \end{tabular}
\end{table}

%%%%%%%%%%%%%%%%%%%%%%%%%%%%%%%%%%%%%%%

\section{Precise statement of the conditions on a chaotic CA}\lb{a-state-cond}

We here present the precise statement of the three conditions on a chaotic CA explained in \sref{analytic}.

\begin{enumerate}
\item We fix a finite $l$, and consider $2(l+1)$ sites in C$_{i,i+l}$.
Let $s_{i,i+l}^n$ be a state of C$_{i,i+l}$ at the $n$-th step.
We require that for any cycle $k$ with no spatial periodicity and for any states of C$_{i,i+l}$ denoted by $y\in (\calS_d)^{2(l+1)}$, the following relation
\eq{
\lim_{L\to \infty} \frac{\sum_{n=1}^{u_k} \chi \( s_{i,i+l}^n=y\) }{u_k}=\frac{1}{d^{2(l+1)}}
}
is satisfied, where $\chi (\cdot )$ takes one if the statement in the clause is true and takes zero otherwise.

\item We require that the maximum length of a cycle is exponentially small compared to the number of possible states:
\eq{
-\lim_{L\to \infty}\frac{1}{L} \ln \frac{\max_k u_k}{d^{2L}}\neq 0.
}

\item We fix a finite $l$, and consider states of $2(L-l+1)$ sites in C$_{i,i+L-l}$ which we denote by $s_{i, i+L-l}^n$.
For any cycle $k$, we construct a subset of $\{ 1,2, \cdots, u_k\}$ as 
\eq{
\calD_k=\set{ n| \exists n'\neq n \ {\rm s.t.} \  s_{i,i+L-l}^n=s_{i,i+L-l}^{n'}}.
}
We then require that the size of $\calD _k$ is negligibly small:
\eq{
\lim_{L\to \infty} \max_k \frac{\abs{\calD _k}}{u_k}=0.
}
\end{enumerate}

%%%%%%%%%%%%%%%%%%%%%%%%%%%%%%%%%%%%%%%
\section{Large effective dimension scenario}\lb{a-Deff}

We here give a precise statement on the fact that the large effective dimension ensures the existence of thermalization, which is discussed in \sref{analytic} and \sref{gen}.
We first fix the precision $\delta>0$.
Let $\sfP_{\rm neq}^\delta$ be a projection operator onto the nonequilibrium subspace where there is a macroscopic observable whose density is different from the corresponding microcanonical average by more than $\delta$.
Ordinal thermodynamic system satisfies
\eqa{
\Tr \left[ \sfP_{\rm neq}^\delta \rho^{\rm mc}\right] \leq e^{-\gamma (\delta) L}
}{gamma}
for any large $L$, which exhibits the large deviation property of the microcanonical ensemble.
Here,  $\gamma (\delta)$ is independent of $L$, and it converges to zero as $\delta \to 0$.
A state $\ket{\Psi}$ thermalizes with precision $\delta$ if $\bra{\Psi}\sfP_{\rm neq}^\delta\ket{\Psi}$ converges to zero in the thermodynamic limit.

It is shown that thermalization with precision $\delta$ indeed occurs if the effective dimension of the initial state satisfies~\cite{Tas16}
\eq{
\deff \geq e^{-\gamma (\delta) L} D
}
with $\gamma (\delta)$ given in \eref{gamma}.
Since we should adopt the case of the perfect precision limit $\delta \to 0$ as an ideal limit, the above result can be interpreted as that thermalization is confirmed if $\deff/D$ decays slower than any exponential function of $L$ in the thermodynamic limit.

\end{document}